\documentclass{article}

\usepackage{PRIMEarxiv}

\usepackage[utf8]{inputenc} 
\usepackage[T1]{fontenc}    
\usepackage{hyperref}       
\usepackage{url}            
\usepackage{booktabs}       
\usepackage{amsfonts}       
\usepackage{nicefrac}       
\usepackage{microtype}      
\usepackage{lipsum}
\usepackage{fancyhdr}       
\usepackage{graphicx}       
\graphicspath{{media/}}     

\title{A Brief Yet In-Depth Survey of Deep Learning-Based Image Watermarking
}

\author{
  Xin Zhong, Arjon Das, Fahad Alrasheedi and Abdullah Tanvir \\
  Department of Computer Science \\
  University of Nebraska Omaha \\
  Omaha, USA\\
  \texttt{\{xzhong, arjondas, falrasheedi, atanvir\}@unomaha.edu} \\
}

\usepackage{float}
\usepackage{array}
\usepackage{tabularx}
\usepackage{color}
\usepackage{xspace}

\newcommand\etal[1]{{\textit{et al}. \xspace}}

\begin{document}
\maketitle

\begin{abstract}
This paper presents a comprehensive survey on deep learning-based image watermarking, a technique that entails the invisible embedding and extraction of watermarks within a cover image, aiming to offer a seamless blend of robustness and adaptability. 
We navigate the complex landscape of this interdisciplinary domain, linking historical foundations, current innovations, and prospective developments. 
Unlike existing literature, our study concentrates exclusively on image watermarking with deep learning, delivering an in-depth, yet brief analysis enriched by three fundamental contributions. 
First, we introduce a refined categorization, segmenting the field into Embedder-Extractor, Deep Networks as a Feature Transformation, and Hybrid Methods. 
This taxonomy, inspired by the varied roles of deep learning across studies, is designed to infuse clarity, offering readers technical insights and directional guidance. 
Second, our exploration dives into representative methodologies, encapsulating the diverse research directions and inherent challenges within each category to provide a consolidated perspective. 
Lastly, we venture beyond established boundaries to outline emerging frontiers, offering a detailed insight into prospective research avenues.
\end{abstract}


\section{Introduction}
\label{sec:intro}
Each year, the Internet serves as a conduit for the upload, transfer, and sharing of billions of digital images~\cite{NumPhoto}. 
The advent of sophisticated digital technologies has facilitated the effortless editing, dissemination, and reproduction of images, precipitating a surge in unauthorized usage and concomitant infringement of intellectual property rights of original creators. 
In this context, digital image protection emerges as a critical mechanism to enforce the sanctity of content creators' intellectual property. 
Digital images are not merely visual content; they constitute significant assets for individuals, corporate entities, and various organizations. 
The integrity of these assets is often threatened by unauthorized utilization and replication, a scenario that could culminate in substantial financial deficits and damage to reputation. 
Moreover, the illicit use of personal images can inflict emotional distress, exacerbated when images or videos are circulated devoid of the owner’s acquiescence. 

Digital image watermarking has emerged as a preeminent technique in the domain of image protection, garnering widespread application and acclaim. 
Central to this technique is the covert embedding of informational elements, ranging from logos to copyright notices, directly into the visual content. 
This surreptitious integration ensures that only individuals endowed with the appropriate authorization can extract the watermark, maintaining the confidentiality and integrity of the embedded data. 
Watermarks serve a multifaceted purpose. 
They act as an indelible signature, affirming ownership and bolstering copyright protection by dissuading unauthorized replication and distribution. 
They stand as testaments to authenticity, facilitating the licensing and tracking of image utilization. 
Moreover, they serve as conduits for covert communication, encapsulating hidden messages seamlessly woven into the visual content. 
Lastly, they function as intrinsic detectors, revealing alterations and tampering, thereby upholding the integrity of the original content. 
The versatility of digital watermarking extends its applicability across a plethora of fields. 
It has become an instrumental tool in forensic analyses, enhancing the traceability and verification of digital content. 
In the burgeoning spheres of 5G communication and the Internet of Things (IoT)~\cite{mastorakis2021dlwiot}, watermarking is pivotal in fortifying security and enhancing data integrity. 
Smart cities, characterized by their intricate networks of interconnected digital systems, leverage watermarking to safeguard data and ensure the seamless, secure interchange of information~\cite{amrit2022survey}.

Image watermarking and image steganography are closely related fields, yet with distinct technical and application-specific differences. Both areas explore the intricate process of subtly embedding data within images, ensuring the modifications remain unnoticeable to the unaided eye. However, the differing goals they aim to accomplish lead to distinct technical focuses. 
Image steganography primarily aims to provide a covert channel for information transmission, avoiding detection by unauthorized parties. It hinges on the principles of unpredictability and high payload capacity~\cite{zhu2018hidden, baluja2017hiding}. The former emphasizes resistance against steganalysis techniques, while the latter denotes the ability to embed a significant amount of data without affecting the perceptual quality of the cover image. 
On the other hand, image watermarking seeks to protect the integrity of both the cover image and the embedded watermark. The fundamental aspect of this field is robustness, which refers to the enduring readability of the watermark amidst various potential attacks. This characteristic is crucial, although there are situations where fragile watermarking is preferable and necessary, especially in scenarios like medical imaging where maintaining the original quality of the image is vital~\cite{shih2016high}.

Traditional image watermarking approaches are predominantly characterized by handcrafted embedding and extraction mechanisms. 
These processes often entail the intricate utilization of prior knowledge and a substantial level of expertise in the domain of image processing. 
The inherent dependency on prior information culminates in designs that are tailored for specific cases and exhibit a marked lack of adaptability~\cite{shih2017digital}. 
Such designs are characterized by a uniform application of watermarking patterns across a myriad of images, neglecting the unique content characteristics and quality attributes inherent to each individual image.
In the realm of robustness, a significant limitation manifests. 
Each handcrafted method accentuates a specific attribute or set of attributes, resulting in a fragmented and isolated approach to enhancing watermark robustness. 
The absence of a comprehensive strategy that encapsulates a broad spectrum of potential attacks is conspicuously evident. 
For instance, the watermarking technique premised on quantization index modulation is primarily tailored to counter JPEG compression artefacts~\cite{chen2001quantization}. Concurrently, methods founded upon the principles of log-polar coordinates are intricately designed to mitigate the impacts of rotational manipulations~\cite{kang2010efficient}.
This scenario illuminates an overarching challenge - the lack of a holistic, adaptive, and universally applicable watermarking strategy. 
The juxtaposition of these isolated techniques against the dynamic and multifaceted nature of digital media manipulation threats underscores a significant vulnerability. 
The evolution of manipulative attacks, characterized by their increasing sophistication, demands a parallel evolution in watermarking techniques that is anchored in adaptability, comprehensive threat mitigation, and contextual applicability. 

Deep learning is characterized by algorithms inspired by the structure and function of the brain, known as artificial neural networks. 
These networks are adept at learning from large volumes of data, enabling the extraction of complex patterns and representations. 
Deep learning has catalyzed significant advancements in various fields, including image and speech recognition, natural language processing, and autonomous systems. 
The depth of the networks, characterized by multiple layers of interconnected nodes, contributes to their capacity to perform intricate computations, offering superior performance and predictive accuracy in diverse applications. 
Each layer transforms its input data to increasingly abstract and complex representations, enabling nuanced decision-making and predictions.

In the quest for enhanced robustness and adaptability in image watermarking, deep learning emerges as a formidable ally. 
Unlike their traditional counterparts, deep learning-based watermarking algorithms harbor the potential to learn and adapt~\cite{zhu2018hidden}. 
They encapsulate the capacity to intuitively morph in response to the unique attributes of each image and the evolving landscape of threats. 
This adaptability heralds a new epoch in watermarking - one characterized by enhanced robustness, imperceptibility, and the nuanced balancing of these cardinal attributes. 
As we traverse this trajectory, the integration of deep learning in watermarking is not just an incremental enhancement but a paradigmatic shift. 
It propels watermarking from a static, isolated, and case-specific discipline into a dynamic, adaptive, and holistic domain. 
This evolution is not only pivotal for the enhanced protection of digital media assets but also instrumental in the nuanced balancing of imperceptibility and robustness, ensuring that the integrity and aesthetic value of the digital media are meticulously preserved. 

The necessity of a survey focusing on deep learning-based image watermarking emanates from the rapid advancements and complexities ingrained in this burgeoning field. 
As the integration of deep learning in image watermarking has emerged as a pivotal focus, there is a requisite for a comprehensive, synthesized, and analytical review of the existing literature and methodologies. 
To this end, this paper presents a comprehensive survey of cutting-edge deep learning-based image watermarking techniques, which serves as a reference for the state-of-the-art in deep learning-based image watermarking, summarizing key research directions and envisioning future studies in the domain.

\subsection{Objectives and Distinctiveness of this Survey}
We illustrate the objectives and distinctiveness of our survey by summarizing an overview of the topic concentrations of existing related survey papers in Table~\ref{tab: existing_surveys}. 
Current surveys predominantly orient towards deep learning model architectures, diversified artificial intelligence methodologies, data hiding techniques, and prominent proposals. 
It should be noted that our review is intricately tailored to encapsulate the synopsis of works germane to deep learning-based image watermarking. Consequently, extended domains, including the watermarking of the deep learning models themselves~\cite{li2021survey}, fall beyond the scope of our discussion.

\begin{table}[H] 
\caption{Summary of existing related surveys.\label{tab: existing_surveys}}
\newcolumntype{C}{>{\centering\arraybackslash}X}
\begin{tabularx}{\linewidth}{
>{\hsize=.6\hsize}X  
>{\hsize=1.4\hsize}X 
}
\toprule
\textbf{Methods}	& \textbf{Concentration}\\
\midrule
Gupta and Kishore~\cite{gupta2021survey}	& Summarizing various convolutional neural network model architectures used in deep learning-based image watermarking\\
\midrule
Amrit and Singh~\cite{amrit2022survey}	& Summarizing watermarking using artificial intelligence, machine learning, and deep learning\\
\midrule
Zhang~\textit{et al.}~\cite{zhang2021brief}	&  Reviewing deep learning-based data hiding, classifying based on capacity, security, and robustness, and outlining three commonly used architectures\\
\midrule
Byrnes~\textit{et al.}~\cite{byrnes2021data}	&  Surveying deep learning techniques for data hiding in watermarking and steganography, and categorizing them based on model architectures and noise injection methods\\
\midrule
Singh~\textit{et al.}~\cite{singh2023comprehensive}	&  Reviewing the popular deep-learning model-based digital watermarking methods and summarizing/comparing contributions in the literature\\
\bottomrule
\end{tabularx}
\end{table}

In contrast to existing work, our survey focuses on deep learning-based image watermarking and provides a brief yet in-depth analysis, distinguished by three primary advantages. 
(1) We systematically categorize deep learning-based image watermarking into Embedder-Extractor, Deep Networks as a Feature Transformation, and Hybrid Methods. 
This categorization, grounded in the distinct roles deep learning assumes in various studies, aspires to offer technical insights and guidance.
(2) We study representative methodologies and encapsulate the directions and challenges of research within each specified category, offering a coherent synopsis. 
(3) We extend our discussion to encompass a detailed exploration of prospective research avenues, delineating emerging frontiers in the domain of deep learning-based image watermarking. 

Through the systematic analysis, critical research direction discussion, and prospective outlooks, one primary objective of our survey is to connect past research, present innovations, and future prospects, potentially propelling the field towards refined methodologies, enhanced effectiveness, and broader applicative horizons. 
The rest of this paper is structured as follows: Section~\ref{sec:preliminary} talks about the relevant preliminaries in conventional image watermarking, Section~\ref{sec:category1} categorizes the techniques and provides a survey of image watermarking based on deep learning, Section~\ref{sec:challenge_future} explores potential research avenues for the future, and Section~\ref{sec:Conclusion} draws a conclusion. 

\section{Preliminaries}
\label{sec:preliminary}

\subsection{Traditional Image Watermarking Components}
\label{sec: traditional_wm}

\begin{figure}[!h]%
  \centering
  \vspace{-1.0em}
  \includegraphics[width=0.6\textwidth]{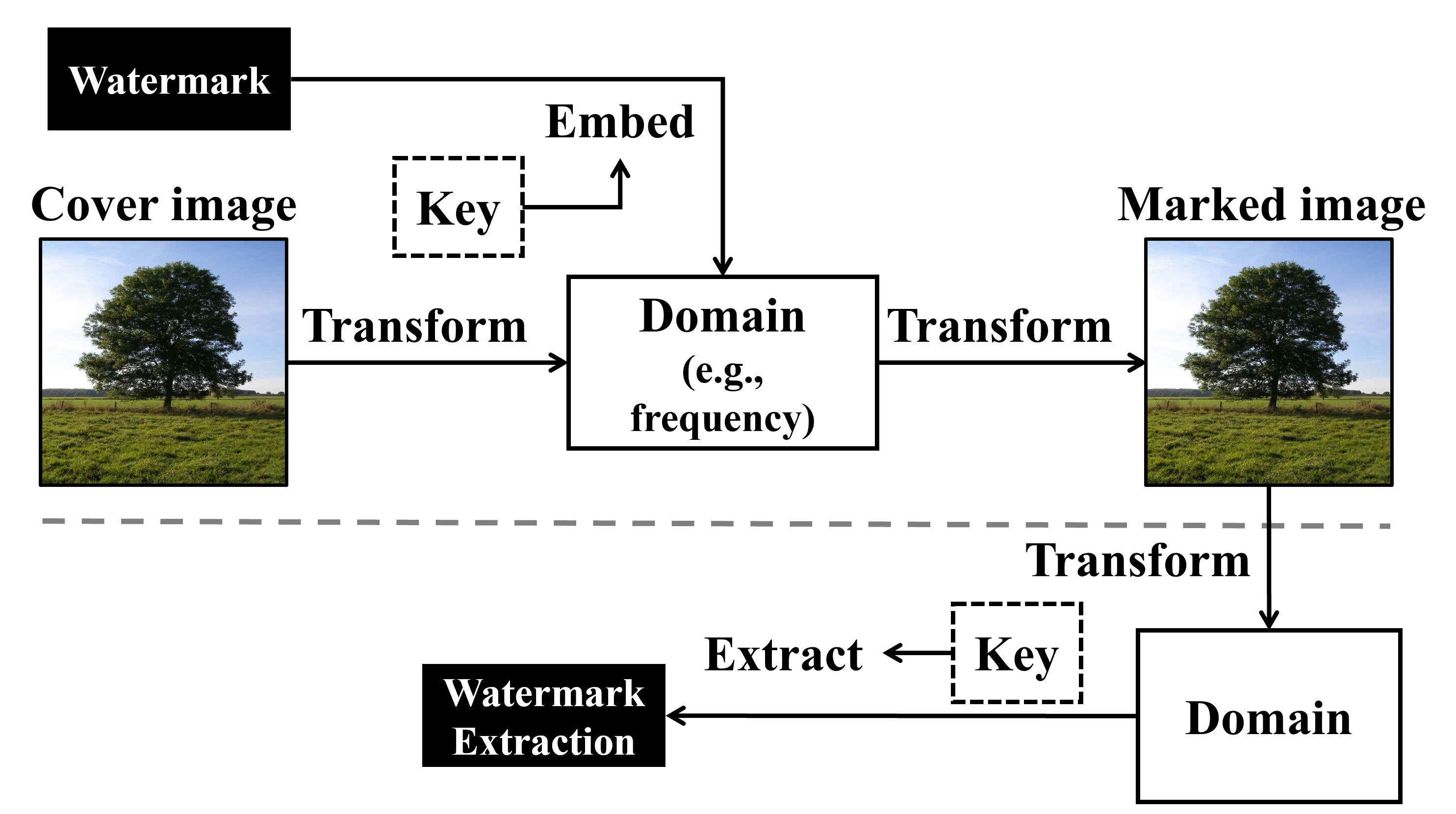}
  \caption{General components and steps of traditional image watermarking.}
  \vspace{-0.5em}
  \label{fig:traditional}
\end{figure}

Image watermarking entails the incorporation of watermark data within an image. 
This watermark, an encoded digital signal, is meticulously crafted to be inconspicuous to human vision yet readily identifiable and extractable via computational algorithms. 
As elaborated in Section~\ref{sec:intro}, the application spectrum of image watermarking, delineated by the nature of the watermark information, spans copyright protection, authenticity verification, covert communication, and tampering detection, among others. 
Fig.~\ref{fig:traditional} succinctly encapsulates the components and steps inherent in the traditional paradigm of image watermarking.

\noindent \textbf{Embedding and Extraction.}
In the embedding step, the watermark is integrated into the cover image via a watermarking algorithm. The overarching objective is to ensure the embedded watermark is robust, resistant to removal or alteration, whilst concurrently maintaining the visual integrity of the cover image. 
Various techniques exist for the watermark embedding, such as the modification of pixel values in the spatial domain~\cite{shih2017digital}, and the manipulation of coefficients within the frequency domain representation~\cite{cox1997secure}. 
Post-embedding, the watermarked image is disseminated to the designated audience, potentially via online platforms. 
Authorized recipients are equipped to extract the embedded watermark utilizing a specialized extraction algorithm. 







\noindent \textbf{Key.} 
Numerous image watermarking methodologies incorporate a key, denoting secret values instrumental in modulating the embedding and extraction processes of the watermark. 
Typically, this key is conjointly generated and disseminated among the content owner and authorized users. 
Its application within the watermarking procedure varies, contingent on the algorithmic design, predominantly aiming to augment security and robustness. 
For instance, the key can be employed to govern the generation of a pseudo-random sequence integrated into the image for watermark embedding~\cite{cox1997secure}, or to designate the precise location of the embedded watermark within the image~\cite{shih2017digital}.

\noindent \textbf{Watermark Preprocessing.} 
Watermark preprocessing can be instrumental in augmenting both security and robustness. 
One classic approach involves the encryption of the watermark using cryptographic techniques, such as AES~\cite{selent2010advanced} or RSA~\cite{zhou2011research}. 
In this process, an encryption key is employed to convert the watermark into a ciphertext, enhancing the security of the embedded data. 
The decryption, and hence the accessibility of the watermark, is contingent upon the application of the corresponding cryptographic key, ensuring the watermark remains impervious to unauthorized access and enhancing its applicability in high-security contexts. In addition to security, the watermark can be encoded using methodologies such as the error correction code~\cite{peterson1972error}, facilitating the rectification of errors within the extracted watermark, enhancing its robustness. 
Reed-Solomon code~\cite{wicker1999reed} and convolutional code~\cite{johannesson2015fundamentals} are classic exemplary methodologies that infuse redundancy into the watermark. 
This inclusion of redundant data is strategically orchestrated to ameliorate errors encountered during watermark extraction, bolstering the accuracy and reliability of the extraction process, even amidst distortions. 

\vspace{-0.5em}
\subsection{Typical Metrics and Factors}
\label{sec: metrics}
Although there can be a large number of evaluation metrics and factors considered in image watermarking based on different applications, certain metrics are ubiquitously employed across both traditional and deep learning-based watermarking in the extant literature. 
In this context, we briefly discuss the typical factors of imperceptibility and robustness, which are integral to assessing the efficacy of watermarking techniques.

\noindent \textbf{Imperceptibility.} 
The ability of the watermark to be embedded into the image data in a way that is invisible to human vision is referred to as imperceptibility. 
The imperceptibility helps ensure that the watermark does not interfere with the quality of the image. 
One most frequently applied evaluation metric is the peak signal-to-noise ratio ($PSNR$): 
\vspace{-0.5em}
\begin{equation}
\label{eq: psnr}
    PSNR = 10 \times log_{10}( 
    \frac{max(c)^2}{
    \frac{1}{\mathcal{R} \mathcal{C}} 
    \sum_{i=1} ^{\mathcal{R}} \sum_{j=1} ^{\mathcal{C}} (c_{ij}-m_{ij})^2} 
    ), 
\end{equation}
where $max(c)$ is the largest possible pixel value for the cover image $c$ which is $255$ if we use 8 bits for each grayscale value, and $\mathcal{R}$ and $\mathcal{C}$ denotes the height and width of the images $c$ and $m$. 

Notably, apart from the extensively utilized $PSNR$, the structural similarity index measure ($SSIM$)~\cite{wang2004image} is also commonly employed to assess imperceptibility, incorporating evaluations of luminance, contrast, and structural disparities. 
An essential augmentation to visual imperceptibility is security~\cite{begum2020digital}. 
This entails ensuring that the embedded watermark is not only invisible to the human eye but also resistant to detection through computational analysis, a criterion of paramount importance in secure watermarking applications, exemplified in domains like smart city planning and digital forensics.

\noindent \textbf{Robustness.} 
Robustness characterizes the watermark's resilience, denoting its capacity to be reliably extracted amidst attacks on the watermarked image, such as compression, filtering, or cropping. 
The assessment of robustness involves calculating the disparity between the extracted and original watermark post-attack. 
When the watermark is binary, the bit error rate (BER) serves as a prevalent metric, computed by dividing the number of erroneous bits by the total number of bits embedded. 
In instances where the watermark takes the form of a two-dimensional matrix, its resilience is often assessed via the normalized cross-correlation ($NC$), measuring the similarity between the original watermark $w$ and the extracted watermark $w'$:
\vspace{-0.5em}
\begin{equation}
\label{eq: nc}
    NC = 
    \frac
    {
    \sum_{i=1} ^{\mathcal{H}} 
    \sum_{j=1} ^{\mathcal{L}} (w_{ij} \cdot w'_{ij})
    }
    {
    \sqrt{\sum_{i=1} ^{\mathcal{H}} \sum_{j=1} ^{\mathcal{L}} (w_{ij})^2}
    \sqrt{\sum_{i=1} ^{\mathcal{H}} \sum_{j=1} ^{\mathcal{L}} (w'_{ij})^2}
    }
    , 
\end{equation}
where $\mathcal{H}$ and $\mathcal{L}$ define the height and width of the watermark. 

\noindent \textbf{Blindness.} 
Blind and semi-blind watermarking represent two prominent approaches in the domain of image watermarking. 
Blind watermarking is characterized by its ability to detect and extract watermarks without the necessity of referring to the original cover image nor the original watermark (while a key may be required). 
On the other hand, semi-blind watermarking, while also not requiring the original cover image for watermark extraction, requires the original watermark (and the key) for extraction.  

\noindent \textbf{Capacity.} 
Capacity denotes the upper limit of watermark data that can be embedded into an image, giving rise to two primary types of image watermarking: 
(1) Zero-bit watermarking focusing on detecting the presence of a watermark in an image rather than extracting data. 
(2) Multi-bit watermarking that embeds and extracts a watermark comprised of multiple bits.

A special case in terms of the capacity factor is zero watermarking, where no watermark is directly embedded~\cite{fierro2019robust}. 
Instead, a relationship is established and stored between the original content and the watermark (a master share). In case of disputes or verification, this relationship is used to demonstrate the presence of the watermark. 
This technique proffers advantages, such as eliminating the need to alter the image, thus preserving its original quality. 
Nonetheless, limitations exist, primarily due to its dependency on the uniqueness of image data, rendering it less effective for images with homogeneous content. 
Additionally, zero watermarking serves as a signature for authenticating image data, lacking the provision to incorporate additional embedded data.

\section{A Brief Yet In-Depth Survey of Deep Learning based Image Watermarking}
\label{sec:category1}
This section discusses the integration of deep neural networks in contemporary deep learning-based image watermarking, explaining the adaptation of traditional watermarking processes within this advanced framework. 
The modern deep learning-based image watermarking approaches are architectured akin to their traditional counterparts, encompassing transformation, watermark embedding, and extraction phases. 
For a structured analysis, we classify current deep learning-based techniques into three distinct categories based on the roles of deep learning in various papers: (1) Embedder-Extractor Joint Training, (2) Deep Networks as a Feature Transformation, and (3) Hybrid methods. 
In order to provide a clear and concise overview of each category, Table~\ref{tab:method_overview} acts as a navigational guide. Each category is outlined, with its key features, requirements, and potential drawbacks listed. 

\begin{table}[h]
\caption{Overview of methods in each proposed categories. \label{tab:method_overview}}
\newcolumntype{C}{>{\centering\arraybackslash}X}
\begin{tabularx}{\linewidth}{
>{\hsize=0.44\hsize}X  
>{\hsize=1.56\hsize}X 
}
\toprule
\textbf{Methodology} & \textbf{Key Features / Requirements / Potential Drawbacks} \\
\midrule
Embedder-Extractor Joint Training & 
\textbf{Key Features:} 
Automated watermarking schemes are learned from designated data through joint optimization of watermark embedding and extraction. \\
& \textbf{Requirements:}
A robust dataset for training and an astute selection of noise levels in the noise layer to ensure robustness. \\
& \textbf{Potential Drawbacks:}
Efficacy may wane with inadequate training data or improper noise selection, emphasizing the necessity for a robust dataset and prudent noise level selection. \\
\midrule
Deep Networks as a Feature Transformation &
\textbf{Key Features:}
Employment of deep networks for feature transformation, leveraging pre-trained models for adept feature extraction. \\
& \textbf{Requirements:}
Pre-training of deep networks on tasks related to robustness within the domain, alongside a separate design of embedding and extraction in this feature space.  \\
& \textbf{Potential Drawbacks:}
Robustness may be compromised if the feature transformation efficacy is subpar, and pre-trained networks might not adequately prioritize robustness. \\
\midrule
Hybrid Methods &
\textbf{Key Features:}
Fusion of classical watermarking with deep learning, harnessing strengths from both realms for enhanced watermarking. \\
& \textbf{Requirements:}
Rigorous design and fine-tuning of both traditional watermarking schemes and deep learning models to ensure harmonious operation. \\
& \textbf{Potential Drawbacks:}
Increased design complexity and potential amplification of limitations inherited from both classical and deep learning-based techniques, necessitating a sagacious design strategy. \\
\bottomrule
\end{tabularx}
\end{table}

\subsection{Embedder-Extractor Joint Training Methods}
\label{sec: both_type}
\begin{figure}[!h]%
  \centering
  \vspace{-1.0em}
  \includegraphics[width=0.65\textwidth]{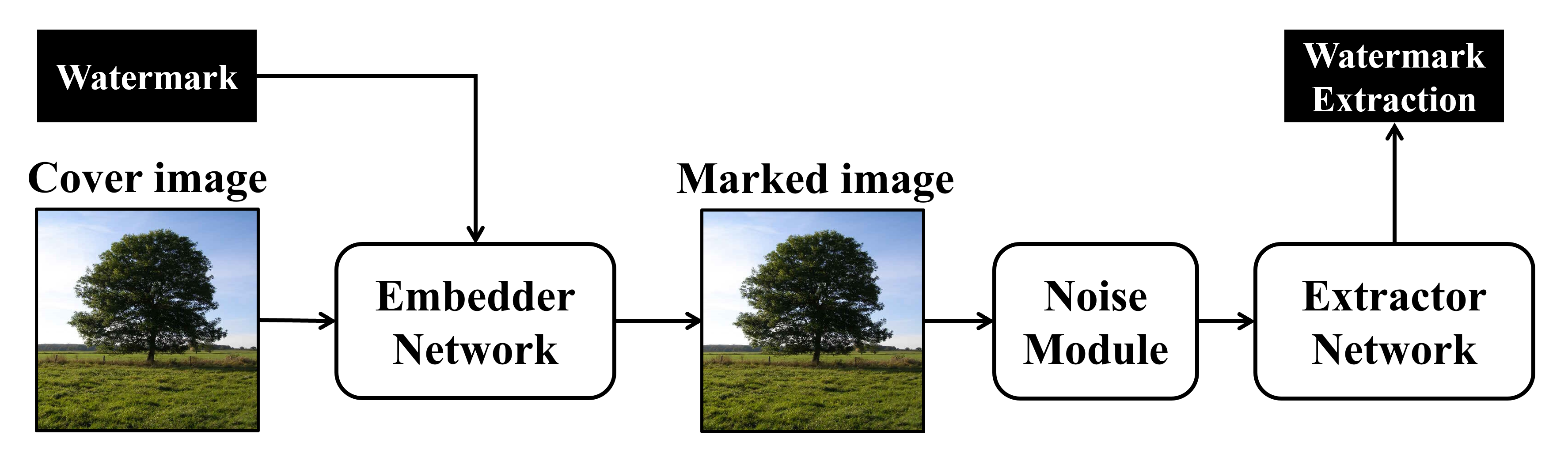}
  \caption{General process of the Embedder-Extraction joint training.}
  \vspace{-0.5em}
  \label{fig:E2E}
\end{figure}

As depicted in Fig.~\ref{fig:E2E}, methods within this category fundamentally involve the training of two core components: an embedder network, responsible for watermark integration into a cover image, and an extractor network, tasked with retrieving the embedded watermark from the marked image. 
Variations in design are present, with some iterations incorporating separate feature extraction networks within the embedder for the preliminary processing of the watermark and cover image. 
To enhance robustness, a noise module is typically positioned subsequent to the marked image. 
This module is instrumental in introducing and amalgamating noise into the marked image during training, equipping the extractor network with an augmented capacity to counteract disturbances.

Typically, all components are jointly trained within a unified deep neural network framework, employing gradient descent as the optimization technique. 
The objective is to optimize a loss function, aiming to uphold the imperceptibility while ensuring the effective extraction of the embedded watermark. 
Mathematically, a representative loss function for such joint training can be defined as:

\vspace{-1.0em}
\begin{equation}
\label{eq: joint_loss}
    \mathit{l} = 
    f_{1}(c, m) + f_{2}(w, w')
    , 
\end{equation}

\noindent where $c, m, w, w'$ represent the cover image, marked image, watermark, and extracted watermark, respectively. The function $f_{1}$ quantifies the visual disparity between $c$ and $m$, aligning with the traditional criterion of achieving a high $PSNR$. Concurrently, $f_{2}$ assesses the variance between $w$ and $w'$, fulfilling the conventional benchmarks of elevated BER or $NC$.

The elements within $\mathit{l}$ present a trade-off: minimizing $f_{1}(c, m)$ to zero implies an exact resemblance between the marked and cover images. 
However, this may constrain the space available for watermark embedding, potentially leading to extraction inefficacy. 
To orchestrate a network capable of harmonizing this trade-off, a prevalent approach involves utilizing the gradients engendered by $f_{2}(w, w')$ to refine the weights of all components, effectuating back-propagation extending to both $c$ and $m$. 
Conversely, gradients emanating from $f_{1}(c, m)$ are exclusively employed to optimize the embedder network’s weights. 
Subsequent sections will unfold the intricate designs and distinct characteristics embodied by representative state-of-the-art methods within this joint training category.

The concept of joint training was first introduced in the HiDDeN paper by Zhu \etal.\cite{zhu2018hidden}. 
The authors aimed to integrate the embedder-extractor paradigm to streamline and unify the processes of image steganography and watermarking. 
Several innovative and practical designs are incorporated in HiDDeN. 
The embedder replicates the watermark, and its integration with the cover image occurs at the embedder’s final layer to maintain the visual quality. 
Discriminator networks are deployed to ascertain the presence of watermarks in images, thereby ensuring that the embedder generates visually coherent marked images.
Robustness is bolstered in HiDDeN through the introduction of a noise layer that incorporates noise-inducing operations, including dropout, Gaussian blur, and JPEG compression during the training phase. 
To navigate the challenges posed by the non-differentiability of the original JPEG  incorporated within the noise layer, HiDDeN introduces a differentiable JPEG variant, ensuring the continuity of the gradient flow. 
Beyond HiDDeN, Zhang \etal.\cite{zhang2020udh} introduced UDH, another seminal work rooted in the joint embedder-extractor paradigm, applicable to both image watermarking and steganography. 
UDH is distinguished as one of the pioneering works to embed entire images as watermarks within embedder-decoder frameworks. 
It introduces an approach to watermark encoding that facilitates disentanglement during the extraction process. 
In this methodology, the encoded watermark is generated independently of the cover image and subsequently integrated, ensuring a systematic and efficient extraction process while preserving the integrity of the cover image.

A potential challenge associated with joint training for robust image watermarking stems from the necessity of the noise layer being differentiable to facilitate gradient flow. 
In addressing this issue, Liu \etal.\cite{liu2019novel} proposed a two-stage training methodology. 
In the initial phase, both the embedder and extractor are collaboratively trained without any noise intervention, ensuring a seamless and undisturbed gradient flow. 
In the second stage, the embedder’s parameters are fixed, rendering it non-adaptable to further training iterations. 
The focus is then channeled exclusively towards the training of the extractor. 
This bifurcated training approach allows the integration of non-differentiable noise layers into the extractor's training without compromising the effectiveness of the whole training process.
This paper has tested its robustness against a spectrum of prevalent attacks, such as resizing, salt \& pepper noise, dropout, crop-out, Gaussian blur, and JPEG compression. 

Numerous endeavors aim to circumvent the challenge posed by the non-differentiable nature of the JPEG operation within the noise layer. 
Chen \etal.\cite{chen2020jsnet} proposed the employment of simulation networks to emulate JPEG lossy compression, accommodating various quality factors. 
The model employs the max-pooling layer, convolution layer, and a noise mask to respectively represent the sampling, DCT, and quantization processes inherent in JPEG compression. 
In a related vein, Jia \etal.\cite{jia2021mbrs} advocated for the incorporation of batches that amalgamate both actual and simulated JPEG compressions. 
Within the training's noise layer, each batch is configured to randomly incorporate either an actual JPEG compression layer, a differentiable simulation of a JPEG layer, or a layer devoid of noise. 
In scenarios employing momentum-based optimization strategies, there's no strict requirement for the joint training of the embedder and extractor. 
However, the embedder is still tailored to generate high quality images robust to JPEG compression, while the extractor is engineered to retrieve features post-JPEG noise. 
Moreover, Zhang \etal.\cite{zhang2021towards} introduced a pseudo-differentiable methodology, designed to accommodate JPEG compression as a specialized noise variant. 
This approach features distinctive forward and backward paths during the training process. Notably, the backward propagation is structured to bypass the JPEG compression phase, thereby mitigating the impediments associated with non-differentiability.

Certain studies incorporate specialized noise into the noise layer to address special challenges in image watermarking. 
One intricate area involves extracting watermarks from images that have undergone resampling via a camera, which introduces multifarious noise types including JPEG artifacts, variations in lighting, and optical distortions. 
In response to this, Fang \etal.\cite{fang2022pimog} and Gu \etal.\cite{gu2023anti} advanced approaches that incorporate a screen-shooting noise layer simulation. 
This adaptation enables the simulation of camera resampling noises like geometric distortions, optical bends, and RGB ripples within the training of deep learning-based image watermarking models, fostering a more robust system capable of counteracting these specific noise introductions.

Existing joint training paradigms necessitate the explicit identification and enumeration of training noise. 
Models tend to exhibit enhanced robustness to noises encountered during training than those not included. 
However, in real-world scenarios, anticipating and listing all potential noises can be impracticable. 
As such, a strand of research is dedicated to forging robust deep learning-based image watermarking models devoid of a prior noise knowledge. 
Zhong \etal.\cite{zhong2020automated} introduced an invariance layer designed to sieve out extraneous information during the watermark extraction phase. 
Within the training ambit, the Frobenius norm of the Jacobian matrix of the invariance layer’s outputs with regard to its inputs is computed and employed as a regularization term. 
The dual objective of minimizing this term, alongside ensuring watermarking requisites (pertaining to the marked image’s quality and watermark extraction efficacy), ensures the output of the invariance layer remains largely invariant to alterations in its input images, hence instilling robustness sans explicit noise enumeration. 
Furthermore, the embedding network employs multi-scale inception networks, facilitating an intricate fusion of the cover image and watermark. 
Another strategy to achieve robustness without resorting to manual noise layer introduction entails the deployment of an adversarial network, serving as an automated assailant. 
Luo \etal.\cite{luo2020distortion} illustrated this by amalgamating an adversarial network within their architecture, functioning as a noise module. 
During training, this adversarial entity, interfaced with the extractor, evolves in proficiency, adept at hampering watermark extraction. 
In counteraction, the extractor strives to mitigate the perturbations induced by the adversarial entity. 
A nuanced calibration of the training process, accentuating the fortification of the extractor against the adversarial network, culminates in a model characterized by enhanced robustness and reliability.

The joint embedder-extractor paradigm stands as a notably effective approach within the existing body of literature. 
Enhanced performance has been a focal point, with innovations in architectural design and training methodologies spearheading advancements. 
Ahmadi \etal.\cite{ahmadi2020redmark} enriched their noise layer with a variety of disturbances including Gaussian, white noise, random cropping, smoothing, and JPEG compression. 
Each training iteration involves a stochastic selection of one specific noise type, ensuring that each assault singularly influences the training loss. 
In another development, Plata \etal.\cite{plata2020robust} unveiled a pre-processor termed a 'propagator', engineered to disseminate the watermark across the image's spatial domain. 
The researchers stratified assorted attacks and corroborated that integrating specific distortions during training augments robustness against an entire category of distortions. 
Echoing the two-stage training approach of Liu \etal.\cite{liu2019novel} and mirroring the noise influences highlighted in HiDDeN~\cite{zhu2018hidden}, Zhang \etal.~\cite{zhang2021blind} introduced a scheme accentuated by a multi-class discriminator connected to the noise-infused marked image. This innovation aims not just at robustness but amplifies the watermark's security within the marked image.
Hao \etal.~\cite{hao2020robust}, while aligning with the visual quality scrutiny embedded in HiDDeN~\cite{zhu2018hidden}, proposed the integration of a high-pass filter at the discriminator’s inception. 
This strategy nudges the watermark into the image’s mid-frequency region, safeguarding visual quality given the amendable nature of high-frequency components. 
The loss computation accords amplified significance to the central region, resonating with the human visual system's focal inclination. 
For noise layer augmentation, incorporations encompass crop, crop-out, Gaussian blur, directional flips, and JPEG compression, painting a comprehensive spectrum of distortions. 
Xu \etal.\cite{xu2021compact} employed a reversible neural network functioning dually as the embedder and extractor, a strategy that is consistent with the traditional, reversible nature of watermarking transformations. 
In a deviation, Mahapatra \etal.\cite{mahapatra2023autoencoder} advocated for the computation and integration of the difference between the marked and cover images into the extractor to augment extraction quality, a technique that transitions away from the blind scheme archetype. 
Zhao \etal.\cite{zhao2022dari} introduced a factor into the embedder, modulating the watermark's intensity on the cover image and employed a trained spatial attention feature map to optimize watermark positioning. 
Ying \etal.\cite{ying2022hiding} targeted an enhancement of embedding capacity, accommodating one to three watermark color images and employing a decoupling and revealing network tandem in the extraction phase. Their noise layer is fortified with cropping, scaling, Gaussian noise, JPEG, and Gaussian blur to simulate realistic distortions.
In a novel structural approach, Fang \etal.\cite{fang2022end} introduced an extractor-embedder-extractor training architecture to bolster extractor efficiency. Their extractor transforms an image into a binary watermark sequence while the embedder crafts an image residual from both the original and decoded watermark, enhancing marked image creation and extraction efficacy. This methodology underscores decoder training, prioritizing extraction quality. 
Incorporating reinforcement learning, Mun \etal.\cite{mun2019finding} enhanced the robustness of the embedder-extractor framework. They amalgamated convolutional neural networks for embedding and extraction and a reinforcement learning actor for noise module operations. The actor, pivotal in selecting and integrating noise types and intensities into marked images, complements the embedder-extractor, which functions as a evaluative environment, assessing the actor's actions and training the extractor to counteract the induced noise efficaciously.

Multimedia has also emerged as a significant focus in the field of deep learning-based image watermarking. 
Das and Zhong~\cite{das2021deep}, for instance, developed a novel method to embed audio watermarks into cover images, accompanied by a network specifically engineered to ascertain the fidelity of the extracted audio watermark to its original form. 
In the domain of document images, Ge \etal.\cite{ge2023robust} proposed a technique enriched with multiple skip connections within the embedder, ensuring the preservation of intricate details in both watermark and cover image. The robustness of the watermarked document images is fortified through the integration of Dropout, crop-out, Gaussian blur, Gaussian noise, resizing, and JPEG techniques in the noise layer. 
Liao \etal.~\cite{liao2021gifmarking} expanded the embedder-extractor watermarking horizon to GIF animations. Their approach entails the employment of three-dimensional deep neural networks, transforming a single watermark into a three-dimensional feature for integration with GIFs. Discriminator networks are employed to evaluate the watermarking efficacy, focusing on maintaining the imperceptibility of the watermarked GIFs, ensuring visual quality while securing embedded data.

Since the introduction of the embedder-extractor joint training concept (a natural extension of traditional watermarking paradigm) with HiDDeN~\cite{zhu2018hidden}, a burgeoning body of literature has meticulously expanded upon this initial idea, giving rise to a diverse array of methodologies. 
As researchers have delved deeper into this domain, an intricate landscape of challenges and corresponding solutions has emerged, each contributing a unique perspective to the overarching narrative of deep learning-based image watermarking. 
These contributions, marked by their innovative approaches to overcoming specific hurdles, underscore the dynamic and evolving nature of this field. 
We have catalogued these varied challenges and solutions, presenting a comprehensive summary in Table~\ref{tab: challenge_embedder_extractor}. 

\begin{table}[H] 
\caption{Summary of the challenges and representative solutions in the Embedder-Extractor image watermarking.\label{tab: challenge_embedder_extractor}}
\newcolumntype{C}{>{\centering\arraybackslash}X}
\begin{tabularx}{\linewidth}{
>{\hsize=0.8\hsize}X  
>{\hsize=1.2\hsize}X 
}
\toprule
\textbf{Challenges}	& \textbf{Representative Solutions}\\
\midrule
The noise layer needs to be differentiable	& Performing a two-stage training scheme\\
\midrule
The non-differentiable nature and low performance issues of JPEG	& Including differentiable JPEG simulations in the noise layer  \\
\midrule
Special challenging noises such as the camera resampling	&  Including simulated camera distortions in the noise layer\\
\midrule
Models have more robustness to trained noises than those not included in the noise layer &  Developing strategies that do not require noise lists during the training, e.g., an invariance layer that sieves out extraneous information, or an adversarial network to automatically attack the extractor\\
\midrule
Aiming at enhanced and improved overall model performance	&  Introducing innovative architectures and training paradigms aligns with the nuanced processes of embedding, extraction, and feature transformation, mirroring the strategic design inherent in conventional watermarking \\
\midrule
Including multimedia for cover images while maintaining high performance	&  Designing special neural networks to process the multi-modal features and robustness \\
\bottomrule
\end{tabularx}
\end{table}

\subsection{Methods Using Deep Networks as a Feature Transformation}
\label{sec: transform_type}
\begin{figure}[!h]%
  \centering
  \vspace{-1.0em}
  \includegraphics[width=0.6\textwidth]{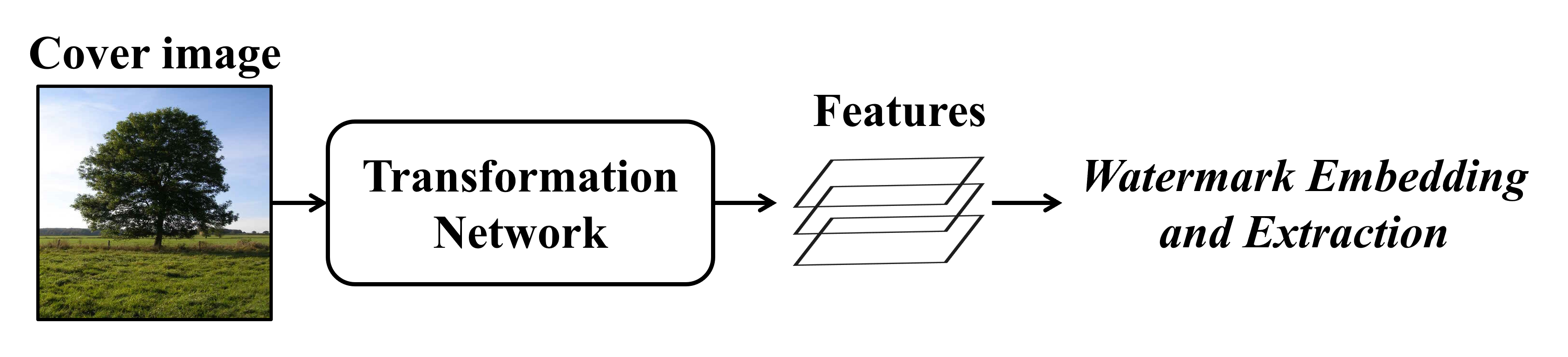}
  \caption{General process of the deep networks as a feature transformation.}
  \vspace{-0.5em}
  \label{fig:trans}
\end{figure}

As illustrated in Fig.~\ref{fig:trans}, watermarking procedures within this category predominantly utilize deep neural networks for feature transformations. 
Both cover and marked images undergo transformations facilitated by these networks, leading to the creation of distinct feature spaces. 
Subsequent watermark embedding and extraction are executed within these defined spaces. 
A prevailing anticipation is the robustness of the transformed domains, implying that even minor alterations to the marked images should yield consistent or nearly identical feature values.

Numerous methods have adopted the concept of Deep Networks as a Feature Transformation in the context of deep learning-based zero watermarking. 
Fierro \etal.\cite{fierro2019robust} employed convolutional neural networks (CNNs) to extract features from cover images, which were then integrated with a permuted binary watermark sequence via an exclusive or (XOR) operation to create a master share. The same CNN processes a test image to extract features, which are subsequently XORed with the master share to extract the watermark. An appropriate key can ensure the identification of the watermark. 
He \etal.\cite{he2023shrinkage} extended this foundational approach by adding fully-connected layers to draw shallow features from various convolutional layers, enhancing the master share creation process. Their introduction of a shrinkage module facilitates the automatic learning of soft thresholding for each feature channel, enhancing feature extraction precision. To optimize the feature space, they focused on eliminating redundancy by learning inter and intra-feature weights and incorporated a noise layer during feature training to increase robustness. 
Han \etal.\cite{han2023application} enhanced this methodology by introducing a chaotic encryption algorithm to encrypt the watermark before the XOR operation, enhancing security. They also adopted the Swin Transformer\cite{liu2021swin} to generate features for master share creation, achieving a feature space that is invariant to geometric distortions and enhances the robustness of the watermarking process.

Another research direction in this category involves employing pre-trained deep neural networks, wherein the training of input data is performed to yield the intended marked images. 
In this scenario, the pre-trained weights remain static, and the alterations in the input are driven towards achieving specific objectives. 
The resultant marked image is visually analogous to the original cover image, yet reveals the embedded watermark upon undergoing feature extraction by the deep network. 
Vukotic \etal.\cite{vukotic2020classification} illustrated this by implementing pre-trained convolutional neural networks and adaptively modifying the input cover image through gradient descent. The dual-faceted loss function encompasses a term that minimizes the perceptual discrepancy between the cover and marked images and another that ensures watermark detectability via a dot product operation, expressed as $\varphi(m)^T\cdot k$. Here, $\varphi(m)$ denotes the marked image's feature extraction through the deep network and $k$ is a predetermined key, facilitating the detection of watermark presence. 
Expanding on this, Fernandez \etal.\cite{fernandez2022watermarking} introduced the capability of multi-bit extraction by assigning distinct keys to each bit of the binary watermark sequence. Contrary to the utilization of convolutional networks pre-trained for classification, they employed networks that had undergone self-supervised learning. This strategic adoption confers a distinct advantage, as the feature spaces derived from self-supervised learning are characterized by augmented robustness, thereby enhancing the effectiveness and reliability of the watermark extraction process.

The utilization of deep networks as a feature transformation represents a nascent avenue in the domain of image watermarking. This approach deviates from the more intuitive Embedder-Extractor joint training model, which seamlessly aligns with traditional image watermarking paradigms by encapsulating both embedding and extraction processes. Consequently, the academic literature on this innovative method is relatively sparse. Nonetheless, this emerging methodology paves the way for captivating research trajectories, offering fresh perspectives and approaches in the field. To provide a consolidated overview, we have collated the prevailing challenges and representative solutions in Table~\ref{tab: challenge_deep_feature_trasnform}. 

\vspace{-0.5em}
\begin{table}[H] 
\caption{Summary of the challenges and representative solutions in the image watermarking using Deep Networks as a Feature Transformation.\label{tab: challenge_deep_feature_trasnform}}
\newcolumntype{C}{>{\centering\arraybackslash}X}
\begin{tabularx}{\linewidth}{
>{\hsize=0.8\hsize}X  
>{\hsize=1.2\hsize}X 
}
\toprule
\textbf{Challenges}	& \textbf{Representative Solutions}\\
\midrule
How to utilize the fitting ability of deep learning to extract the cover image feature (the master share) in zero watermarking & Applying off-the-shelf CNNs or Transformers and designing extended branches of these architectures \\
\midrule
How to choose appropriate deep learning models for image watermarking given their different feature extraction abilities and purposes & Adopting pre-trained CNNs or the models in self-supervised learning   \\
\midrule
How to design separate embedding and extraction schemes given a deep learning feature extractor & To obtain a marked image, fix the pre-trained model, and update the input image with the gradient (produced by a loss ensuring the imperceptibility and extraction integrity) \\
\bottomrule
\end{tabularx}
\end{table}

\subsection{Hybrid Methods}
\label{sec: combine_type}

Methods encompassed in this category exhibit a fusion of deep learning techniques and traditional calculations associated with image watermarking. 
Such an integration implies a symbiotic relationship where the strengths of one approach compensate for the weaknesses of the other, resulting in enhanced efficiency and effectiveness. 
The design paradigms and operational frameworks of these methods can be diverse, exhibiting a wide range of structural and functional variations. 
In these hybrid systems, deep learning typically plays a pivotal role in watermark extraction. The complex and intricate architectures of deep learning models offer enhanced capacity for fitting complex functions, and these models are adept at uncovering intricate patterns and correlations within the watermarked images, thereby facilitating the efficient and accurate extraction of embedded watermarks. 
The conventional image watermarking calculations, on the other hand, lend stability, reliability, and a degree of interpretability to the process. They serve as a solid foundation upon which the deep learning models can build.

Kandi \etal.\cite{kandi2017exploring} employed two convolutional autoencoders to reconstruct a cover image individually. The distinctions between the autoencoder-reconstructed images and the original cover image are integral to their approach. The first autoencoder's reconstruction denotes bit zeros in a binary watermark, while the second represents bit ones. 
In a different context, Ferdowsi \etal.\cite{ferdowsi2018deep} tailored a technique specifically for Internet of Things (IoT) applications, utilizing Classic Spread Spectrum for watermark embedding, wherein a key pseudo-noise sequence augments the original signal. The innovation lies in mapping features like spectral flatness, mean, variance, skewness, and kurtosis of the cover image to bit streams, serving as the watermark, enhancing security against eavesdropping attacks by eschewing predefined bit streams. 
Li \etal.\cite{li2019novel} introduced a method where pre-processed grayscale watermark images are integrated into the DCT blocks of cover images. Extraction of these embedded watermarks is facilitated by training convolutional neural networks, establishing a bridge between conventional and neural approaches. 
Mellimi \etal.\cite{mellimi2021fast} advocated for embedding watermarks into the lifting wavelet domain~\cite{sweldens1995lifting} of cover images. They introduced noise into the marked image and deployed a deep neural network as an extractor, exemplifying robustness to the infused noise. 
Zhu \etal.\cite{zhu2021robust} innovated a technique amalgamating key point detection with deep learning-based image watermarking. Utilizing SURF~\cite{bay2006surf}, they delineated scale-invariant embedding regions, placing normalized binary watermarks at their centers in the Y color channels. This aggregated data is routed through an embedding network, yielding the marked Y channel. An extractor network, fine-tuned through training, facilitates watermark retrieval. Robustness is amplified by the incorporation of perturbations during the training phase, underscoring an enhanced resilience to various forms of distortions and manipulations. 
Chack \etal.\cite{chacko2022deep} introduced a hybrid methodology that intertwines traditional watermarking, convolutional neural networks (CNN), and evolutionary optimization. This multifaceted approach embeds an Arnold-transformed watermark into the DCT domain, employs Harris Hawks optimization to fine-tune the embedding strength, and relies on a CNN to uncover the embedded watermark. 
In a separate study, Fang \etal.\cite{fang2020deep} presented a deep template-based image watermarking mechanism. The embedding process in their approach encodes the watermark using established techniques, employs an auxiliary locating template to manipulate a pseudo-random Gaussian noise pattern, and integrates the watermark into the red and blue color channels of a cover image. Extraction is facilitated by two deep neural networks; the initial network extracts and accentuates features, while the subsequent network classifies the watermark bit patterns. 
Kim \etal.\cite{kim2020convolutional} presented another nuanced technique utilizing templates for watermarking images. Their strategy involves segmenting a cover image into distinct patches, earmarking specific patches for watermark embedding, and others for housing a predefined template. Watermark insertion is executed in the Curvelet domain via quantization index modulation, while the template undergoes processing by a dedicated generation network before integration into the cover image. The marked image is derived by assembling the various embedded patches. Extraction is facilitated by a template extraction network that unveils the embedded template, which is subsequently juxtaposed against the original via a template-matching network. This comparison process facilitates the identification of potential geometric distortions inflicted upon the marked image. 
Chen \etal.\cite{chen2021wmnet} prioritized the development of a mechanism for authenticating watermark systems via deep learning. Specifically, their framework is adept at discerning the accuracy of watermark extractions from medical images. Their innovative approach involves simulating a variety of watermark distortions and compiling a labeled dataset. This dataset then undergoes training on a neural network designed to validate the integrity of extractions derived from potentially marked images, thus bridging the gap between watermark verification and deep learning methodologies.

The integration of deep learning and traditional image watermarking has given rise to a plethora of methodologies, each characterized by its distinct approach and underlying principles. 
Despite the diversity inherent in these hybrid methods, it is noteworthy that they tend to encounter a set of common challenges and have consequently adopted prevailing solutions to mitigate these issues effectively. 
These challenges largely stem from the complex interplay between the adaptive, data-driven nature of deep learning and the algorithmic, rule-based structure of traditional watermarking. 
Addressing these issues necessitates a nuanced approach that is sensitive to the strengths and limitations inherent in both paradigms. 
In Table~\ref{tab: challenge_hybrid}, we have compiled a summary of typical challenges and their corresponding solutions. 

\vspace{-0.5em}
\begin{table}[H] 
\caption{Summary of the challenges and representative solutions in Hybrid Methods.\label{tab: challenge_hybrid}}
\newcolumntype{C}{>{\centering\arraybackslash}X}
\begin{tabularx}{\linewidth}{
>{\hsize=0.8\hsize}X  
>{\hsize=1.2\hsize}X 
}
\toprule
\textbf{Challenges}	& \textbf{Representative Solutions}\\
\midrule
Determining the optimal role of deep learning in hybrid watermarking frameworks & Employing deep learning to enhance watermark extraction processes \\
\midrule
The integration of deep learning and traditional watermarking techniques often results in augmented complexity & Crafting modular and scalable architectures facilitates seamless integration and interoperability between both methodologies   \\
\midrule
Refining the synergy between embedding and extraction processes is essential, given the distinct strengths and weaknesses inherent to each approach & Utilizing deep learning for its adaptability and learning prowess, complemented by leveraging the proven properties of traditional algorithms \\
\bottomrule
\end{tabularx}
\end{table}

\section{Discussion of Potential Future Directions}
\label{sec:challenge_future}
\begin{figure}[!h]%
  \centering
  \vspace{-1.0em}
  \includegraphics[width=0.75\textwidth]{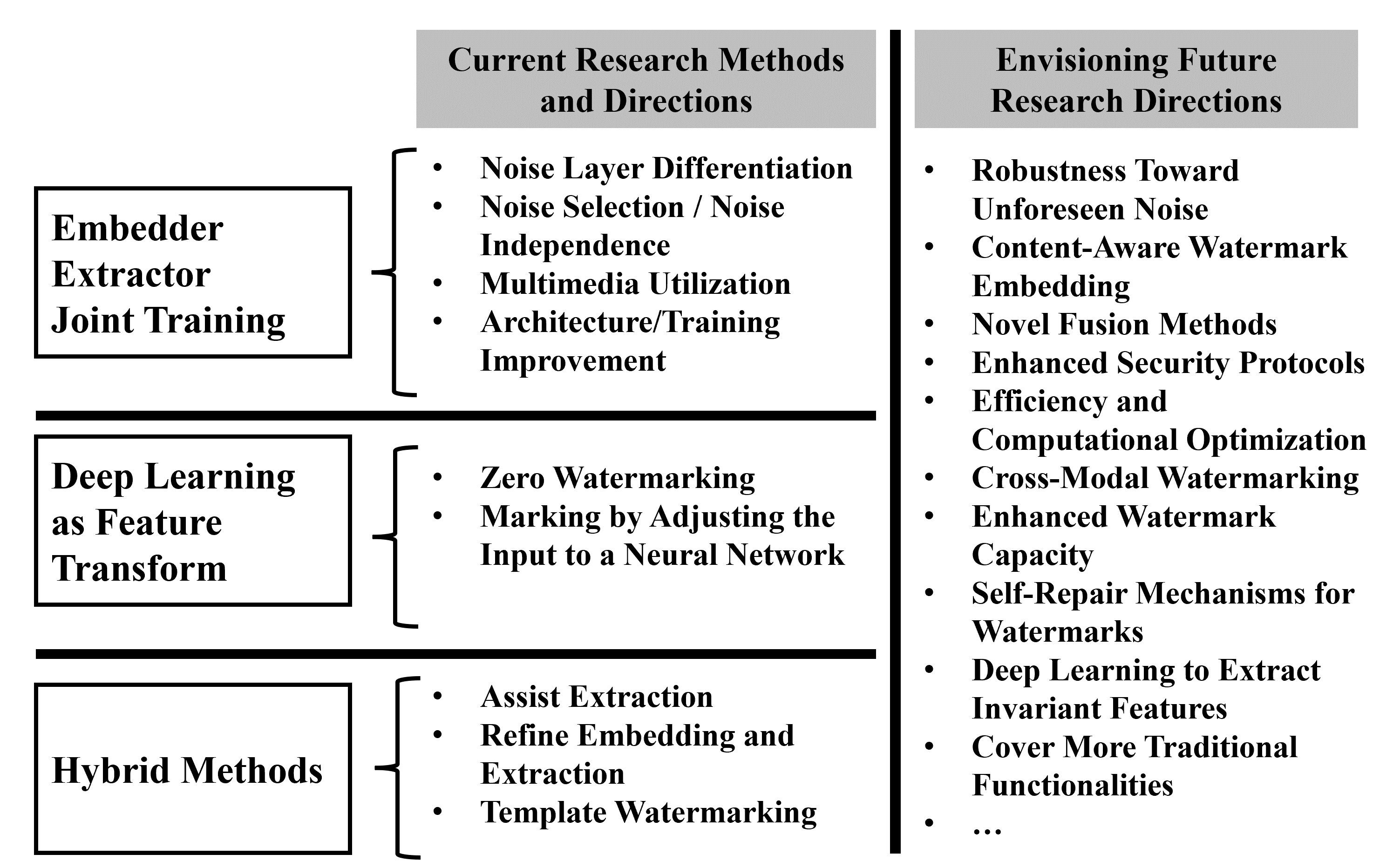}
  \caption{Summarizing and envisioning research directions in deep learning-based image watermarking}
  \vspace{-0.5em}
  \label{fig:summary}
\end{figure}

The proliferation of proposals concerning deep learning-based image watermarking has instigated our comprehensive survey aimed at bridging historical, contemporary, and prospective research. 
Fig.~\ref{fig:summary} encapsulates the prevailing trends delineated in Section~\ref{sec:category1} and extrapolates future investigative trajectories. 
The contemporary focus gravitates towards the intricacies of noise layer differentiation, diversity in noise types, enhancements in architecture and training paradigms, and the strategic integration of deep learning within image watermarking. 
This survey underscores a spectrum of untapped research avenues that transcend traditional frameworks. These emergent perspectives are poised to foster considerable innovations in this domain, skillfully navigating the complex interplay of robustness, imperceptibility, capacity, and security requisite in the dynamic realm of digital media and communications. 
The remainder of this Section discusses our proposals for potential research directions for the future.

\noindent \textbf{Robustness Toward Unforeseen Noise}.
Deep learning-based image watermarking models exhibit distinct robustness to various types of noise, a characteristic intricately linked to the specific noise types they are trained on. 
This variance in robustness is prominently observed when contrasting the model's performance against trained and untrained noise types. Trained noise types are those which the model has been explicitly exposed to during the training phase, enabling the development of specialized mechanisms to mitigate their impact. 
Consequently, the model's efficacy in watermark extraction remains largely stable when encountered with these familiar noise types. 
Conversely, untrained noise types introduce an element of unpredictability. 
Since the model lacks prior exposure and adaptive development against these noise types, its performance can potentially be compromised. 
This differential in robustness underscores the critical importance of a comprehensive training regimen that encompasses a diverse array of noise types to bolster the model's generalization capabilities. 
Future research could focus on enhancing model adaptability and robustness against untrained noise types, perhaps through the integration of adaptive learning mechanisms, online learning methods, or meta-learning strategies that equip the model to swiftly acclimatize to unfamiliar noise environments. 

\noindent \textbf{Content-Aware Watermark Embedding}.
A predominant focus of deep learning-based image watermarking has been accorded to static images. 
This static orientation potentially undermines the watermark’s efficacy, given that optimal embedding strategies can significantly vary across different content types and dynamic scenarios. 
A transition towards content-aware watermark embedding techniques has the potential to redress this imbalance. This approach, conceptualized to be inherently adaptive, is envisaged to utilize sophisticated algorithms capable of analysis and adaptation to the unique attributes of each image or media sequence. 
For instance, CNNs or similar deep learning architectures could be trained to discern intricate patterns and variances in visual content, enabling the model to adapt watermark placement and intensity based on different image regions. 
This would ensure that the watermark is not only imperceptible but also robust against various attacks, establishing a harmonious balance between visibility and security, and marking a significant stride in the advancement of image watermarking technologies.

\noindent \textbf{Novel Fusion Methods}.
Investigating innovative algorithms and techniques for the fusion of watermarks within cover images is crucial. 
A meticulous investigation into innovative embedding strategies is pivotal to ascertaining a harmonious blend that ensures both the visual integrity of the cover image and the indelibility of the watermark. 
Current methods mainly apply additive fusion and concatenation. 
Additive fusion integrates the watermark into the cover image by additive amalgamation, and concatenation involves the direct attachment of watermark features to the cover image. 
Future works can focus on the embedding algorithms to ensure that the watermarks are intricately woven into the cover images, balancing perceptual transparency and robustness against removal or attacks. 
One prospective method can be cross attention which can leverage the attention mechanism to selectively focus on specific features of the cover image during the embedding process, ensuring a dynamic and adaptive incorporation of the watermark.

\noindent \textbf{Enhanced Security Protocols}.
The imperatives of security, privacy, and integrity are being redefined by the sophistication of adversarial attacks. Consequently, the integration of innovative security protocols is not just desirable, but essential. 
A compelling research trajectory could involve the synthesis of cutting-edge cryptographic algorithms with deep learning, an amalgamation promising enhanced watermark protection. 
The incorporation of blockchain technology presents another frontier, offering decentralized, immutable, and transparent platforms for watermarking data transactions and validations. 
These multifaceted, integrative approaches are predicated on a nuanced understanding of both deep learning intricacies and the dynamics of contemporary cryptographic paradigms. As we forge ahead, the synthesis of these technologies could engender a new epoch of resilience, privacy, and security in deep learning-based image watermarking.

\noindent \textbf{Efficiency and Computational Optimization}.
The dynamic landscape of deep learning-based image watermarking is increasingly underscored by the imperative to balance computational efficiency and processing power, especially for real-time applications. 
There lies a complex interplay between ensuring robust watermarking and the computational load, where an optimal middle ground is sought to ensure efficiency without compromising performance. 
In this context, the conceptualization and development of lightweight architectures and algorithms embody a critical focal point of future research trajectories. 
One promising avenue involves the integration of quantization and pruning techniques within the deep image watermarking models, aiming to reduce the model size while preserving the watermarking efficacy. Furthermore, the exploration of knowledge distillation could facilitate the training of compact models that inherit the performance characteristics of larger, more complex models, thereby ensuring efficiency and efficacy in tandem.

\noindent \textbf{Cross-Modal Watermarking}.
In the evolving sphere of deep learning-based image watermarking, cross-modal watermarking emerges as a frontier that offers unprecedented opportunities and challenges. 
It signifies the confluence of diverse media types, extending the watermarking paradigm beyond its traditional confines, and fostering a multi-dimensional approach to content protection and authentication. 
Embedding watermarks in images that can be subsequently extracted from audio or video entails a complex interplay of algorithms and technologies, necessitating innovation and adaptability. 
One methodological prospect could involve the integration of transformer-based models, renowned for their capability to handle varied data types and complexities. Such models can be designed to embed intricate watermark patterns in images, with complementary algorithms tailored for the extraction of these patterns from audio or video formats. A synchronization protocol, ensuring the congruence of embedding and extraction processes across different media, would be integral to this approach.

\noindent \textbf{Self-Repair Mechanisms for Watermarks}.
The integration of self-repair mechanisms in deep learning-based image watermarking presents an avant-garde approach to enhance the robustness and sustainability of watermarks amidst distortions or attacks. 
A watermark endowed with self-repair capabilities can significantly augment the reliability of information authentication and integrity verification processes. 
This concept aligns with the notion of regenerative embedding patterns that maintain their integrity even when subjected to complex distortions or malicious interventions. 
Algorithmically, this could be achieved through the incorporation of redundant encoding schemes, where the critical information is dispersed within the watermark in a manner that allows for reconstruction from partial data. 
Error correction codes and machine learning models adept in pattern recognition and restoration can be synergized to enhance the watermark's robustness. By employing neural networks trained to identify and rectify distortions, the watermark’s intrinsic characteristics can be preserved.

\noindent \textbf{Deep Learning to Extract Invariant Features}.
Current advancements in deep learning-based watermarking leverage pre-trained neural networks, specifically honed through self-supervised learning, to bolster robustness against noise within the transformed domain. 
Such advancements are anchored on the premise that various self-supervised networks, especially those employing joint feature-embedding and contrastive learning methodologies~\cite{chen2020simple, caron2021emerging}, are adept at mitigating the impacts of diverse noise typologies. 
These networks ensure that multiple augmentations of a single image yield identical feature representations. 
Nevertheless, contemporary contrastive learning is predominantly oriented towards evaluating the representational efficacy of the learned space~\cite{da2022solo}. The metric for assessing this efficacy hinges on the network’s performance in tasks encompassing classification, segmentation, and low-shot learning. 
Invariant feature training is an ancillary aspect, not the central focus of these learning paradigms~\cite{bardes2021vicreg}. 
Given this context, the direct application of pre-trained self-supervised neural networks to image watermarking can be impractical, primarily because these networks often neglect to consider ubiquitous distortions for image watermarking like perspective transformations. 
Consequently, there exists a notable research void warranting exploration—formulating specialized self-supervised neural networks expressly tailored for image watermarking applications. These networks would be instrumental in confronting geometric distortions, including but not limited to, rotations and perspective alterations, underscoring a pivotal frontier for ensuing inquiries.

\noindent \textbf{Cover More Traditional Functionalities}.
For deep learning-based image watermarking, a pronounced gap exists in adequately addressing traditional imperatives such as tamper detection. 
Contemporary methodologies predominantly concentrate on robustness, imperceptibility, and capacity, often sidelining the quintessential aspect of detecting alterations or manipulations in the watermarked images. 
Classical watermarking techniques have showcased efficacy in this domain, enabling the identification of unauthorized modifications with reasonable accuracy. 
Incorporating advanced deep learning architectures could potentially elevate the precision and reliability of tamper detection. 
One plausible approach involves the integration of CNNs trained to discern subtle alterations in the watermarked images, leveraging their capacity for feature extraction and pattern recognition. Another avenue could be the exploration of recurrent neural networks (RNNs) to analyze sequences of image data for temporal alterations, offering insights into the progression of tampering efforts.

The primary focus of this paper resides in a comprehensive examination of image watermarking with deep learning, yet it is acknowledged that there exists a spectrum of compelling research areas that, albeit unexplored in this treatise, hold significant relevance and intrigue. 
Instances of such topics include the embedding of watermarks within deep learning models to bolster their protection, as highlighted by Uchida \etal.\cite{uchida2017embedding}. 
Another noteworthy area is the study of watermarking neural networks with watermarked images as input, explored in the work of Wu \etal.\cite{wu2020watermarking}. 
Additionally, the intriguing domain of launching attacks on neural networks utilizing watermarked images is presented in the research by Jiang \etal.~\cite{jiang2021fawa}. 
Each of these areas presents a rich vein of inquiry that complements the broader landscape of image watermarking research.

\section{Conclusion}\label{sec:Conclusion}
In this paper, we have delved into the nuanced realm of image watermarking, a technique characterized by the subtle integration and retrieval of watermarks within a cover image. 
The motivation for this investigation is spurred by the growing synergy between image watermarking and deep learning—a field renowned for its adeptness at unraveling intricate patterns and representations. 
This study stands as a comprehensive exploration, not merely retracing the trajectories of extant methodologies but weaving the tapestry that connects the legacy, current innovations, and future prospects of deep learning in image watermarking.

Distinctive in its approach, this survey illuminates the landscape of deep learning-based image watermarking, marked by its precision and depth of analysis. 
It offers three primary contributions to the scholarly discourse. 
First, we introduce a systematic classification that segments deep learning-based image watermarking into three core categories: Embedder-Extractor, Deep Networks as a Feature Transformation, and Hybrid Methods. 
This refined categorization is premised on the diverse roles that deep learning occupies in related studies and is crafted to infuse clarity and direction into the ongoing research. 
Secondly, we examine emblematic methodologies and encapsulate the multifaceted directions and challenges that each category embodies. 
This aims to provide readers with a consolidated, insightful overview, distilled from a plethora of diverse yet interconnected researches. 
Finally, our analysis expands to unravel prospective research trajectories, mapping out uncharted territories and emergent themes in the field of deep learning-based image watermarking.

\bibliographystyle{unsrt} 
\bibliography{references}

\end{document}